# Speed of sound data, derived perfect-gas heat capacities, and acoustic virial coefficients of a calibration standard natural gas mixture and its $H_2$-enriched blend

Daniel Lozano-Martín[a], David Vega-Maza[a], Alejandro Moreau[a], M. Carmen Martín[a], Dirk Tuma[b], José J. Segovia[a,]*

[a] TERMOCAL Research Group, Research Institute on Bioeconomy (BioEcoUVa), University of Valladolid, Escuela de Ingenierias Industriales, Paseo del Cauce 59, 47011 Valladolid, Spain.

[b] BAM Bundesanstalt für Materialforschung und -prüfung, D-12200 Berlin, Germany.

* Corresponding author e-mail: jose.segovia@eii.uva.es

**Abstract**

This work aims to address the technical aspects related to the thermodynamic characterization of natural gas mixtures blended with hydrogen for the introduction of alternative energy sources within the Power-to-Gas framework. For that purpose, new experimental speed of sound data are presented in the pressure range between (0.1 up to 13) MPa and at temperatures of (260, 273.16, 300, 325, and 350) K for two mixtures qualified as primary calibration standards: a 11 component synthetic natural gas mixture (11 M), and another low-calorific $H_2$-enriched natural gas mixture with a nominal molar percentage $x_{\mathrm{H_2}} = 3$ %. Measurements have been gathered using a spherical acoustic resonator with an experimental expanded ($k$ = 2) uncertainty better than 200 parts in $10^6$ (0.02 %) in the speed of sound. The heat capacity ratio as perfect-gas $\gamma^{\mathrm{pg}}$, the molar heat capacity as perfect-gas $C_{\mathrm{p,m}}{}^{\mathrm{pg}}$, and the second $\beta_{\mathrm{a}}$ and third $\gamma_{\mathrm{a}}$ acoustic virial coefficients are derived from the speed of sound values. All the results are compared with the reference mixture models for natural gas-like mixtures, the AGA8-DC92 EoS and the GERG-2008 EoS, with special attention to the impact of hydrogen on those properties. Data are found to be mostly consistent within the model uncertainty in the 11 M synthetic mixture as expected, but in the limit of the model uncertainty at the highest measuring pressures for the hydrogen-enriched mixture.

**Keywords:** speed of sound, acoustic resonance, hydrogen, natural gas, heat capacities as perfect-gas, acoustic virial coefficients

## 1. Introduction.

The quest for a sustainable and carbon dioxide-free energy new economy paradigm has become a priority. One of the proposals with promising potential for a cost-effective transition from our current highly energy-dependent model is the, so called, Power-to-Gas [1,2]. The essence of the Power-to-Gas technique is the storage and transport of energy in the form of pressurized hydrogen by blending it with the natural gas so that the existing natural gas network can be used without the need for a separate infrastructure [3]. Provided that hydrogen is produced by: (a) electrolysers powered with the surplus of electric energy from renewable sources, like wind, solar, hydraulic or nuclear plants; (b) the steam reforming of natural gas or gasification and reforming of coal, oil, and biomass with carbon capture utilization and storage technologies, there is a reduction in net carbon dioxide emissions [4]. Hydrogen can be used either pure or blended with natural gas as a fuel, as feedstock for methanation [5], or for production of chemical and other fuels production after separation.

Implementing Power-to-Gas projects requires models that correctly describe the thermodynamic behavior of natural gas mixtures enriched with hydrogen. The aim of this research is to improve the accuracy and assess uncertainties of the equations of state that model the thermodynamic behavior of those mixtures, by virtue of the discussion of the differences between the experimental speed of sound and its derived heat capacity and the models' predictions on a 11 component synthetic natural gas (11 M) and a $H_2$-enriched mixture for wide ranges of temperature and pressure. Additionally, accurate measurements of the speed of sound are of interest in the monitoring of composition changes in a gas line [6,7], the design of pipelines to prevent propagation of fractures after decompression [8,9], the calibration of sonic nozzles for gas flow metering [10], or even the implementation of supersonic separators [11,12].

Although there is extensive literature about the speed of sound in multicomponent natural gas-like mixtures [13–18], none of these investigations deals with mixtures containing hydrogen, the objective of this research. Following a previous work reporting accurate experimental (*p, ρ, T*) data with exactly these two gas mixtures [19], we have measured speeds of sound and compared them with the corresponding calculations obtained from the reference thermodynamic mixture models most widely used in the industry, namely the AGA8-DC92 [20,21] and GERG-2008 [22,23], respectively. To accomplish this task, measurements were carried out in the gas and supercritical (*p*,*T*) states depicted in Figure 1, extending at pressures up to 13 MPa at five temperature between 260 K and 350 K. Speed of sound was determined using the most precise experimental technique, the spherical acoustic resonator [24].

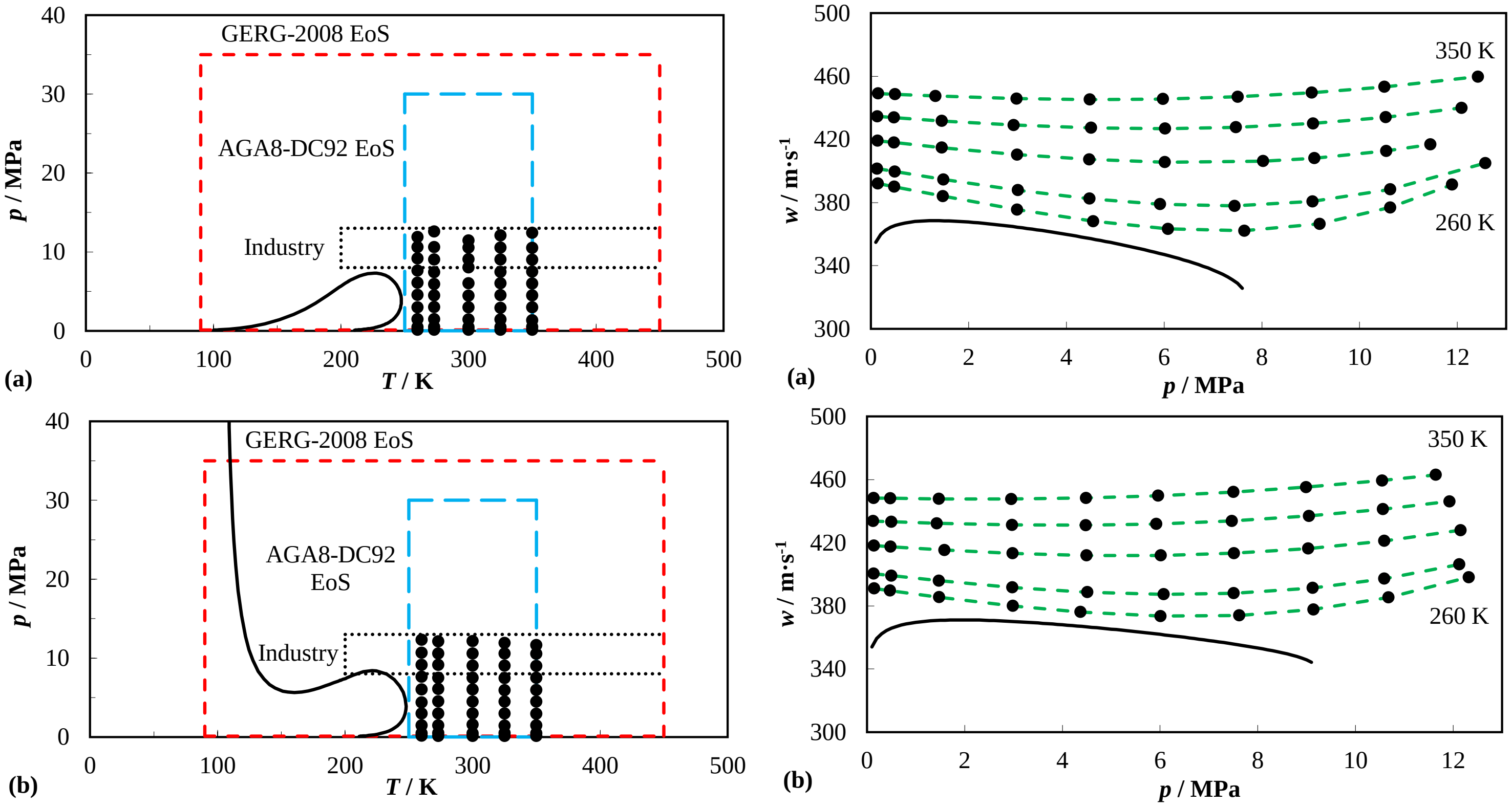

**Figure 1.** (*p*, *T*) and (*w*, *p*) phase diagrams showing the experimental points measured (●) and the calculated phase envelope (solid line) using the GERG-2008 EoS [22,23] for: a) the 11 component synthetic natural gas mixture (11 M) and b) the $H_2$-enriched natural gas mixture. The marked temperature and pressure ranges represent the normal range of application of the AGA8-DC92 EoS [20,21] (blue long dashed line) and the GERG-2008 EoS [22,23] (red short dashed line), respectively, and the area of interest for the gas industry (black dotted line). The green dashed lines serve as a guide for the eye.

## 2. Materials and methods.

### 2.1 Mixtures.

The two gas mixtures studied in this work were prepared by the Federal Institute for Materials Research and Testing (Bundesanstalt für Materialforschung und -prüfung, BAM) in Germany using pure gases with the specifications given in Table 1.

**Table 1.** Purity, supplier, molar mass, and critical parameters of the pure components used for the preparation of the two gas mixtures at BAM.

| | Supplier | CAS-number | Purity / mol-% | $M$ / g·mol$^{-1}$ | Critical parameters[a] | |
|---|---|---|---|---|---|---|
| | | | | | $T_c$ / K | $p_c$ / MPa |
| Methane | Linde AG | 74-82-8 | ≥ 99.9995 | 16.043 | 190.564 | 4.5992 |
| Ethane | Matheson Tri-Gas | 74-84-0 | ≥ 99.999 | 30.069 | 305.322 | 4.8722 |
| Propane | Scott Specialty Gases BV | 74-98-6 | ≥ 99.999 | 44.096 | 369.890 | 4.2512 |
| Butane | Scott UK | 106-97-8 | ≥ 99.95 | 58.122 | 425.125 | 3.7960 |
| Isobutane | Scott Specialty Gases | 75-28-5 | ≥ 99.98 | 58.122 | 407.810 | 3.6290 |
| Pentane | Sigma-Aldrich Chemie | 109-66-0 | ≥ 99.7 | 72.149 | 469.700 | 3.3675 |
| Isopentane | Sigma-Aldrich Chemie | 78-78-4 | ≥ 99.7 | 72.149 | 460.350 | 3.3780 |
| Neopentane | Linde AG | 463-82-1 | ≥ 99.0 | 72.149 | 433.740 | 3.1960 |
| Hexane | Sigma-Aldrich Chemie | 110-54-3 | ≥ 99.7 | 86.175 | 507.820 | 3.0441 |
| Carbon dioxide | Air Liquide AG | 124-38-9 | ≥ 99.9995 | 44.010 | 304.128 | 7.3773 |
| Nitrogen | Linde AG | 7727-37-9 | ≥ 99.9995 | 28.014 | 126.192 | 3.3958 |
| Oxygen | Westfalen AG | 7782-44-7 | ≥ 99.9999 | 31.999 | 154.581 | 5.0430 |
| Helium | Linde AG | 7440-59-7 | ≥ 99.9995 | 4.003 | 5.195 | 0.2283 |
| Hydrogen | Linde AG | 1333-74-0 | ≥ 99.9999 | 2.016 | 33.145 | 1.2964 |

[(a)] Critical parameters were obtained by using the default equation for each substance in REFPROP 10.0 software [25].

The first mixture matches a pipeline-quality rich natural gas composed of 11 components and denoted as 11 M synthetic natural gas, whereas the second mixture resembles a low-calorific 12 components natural gas blended with hydrogen denoted as $H_2$-enriched natural gas. The normalized compositions and corresponding expanded ($k$ = 2) uncertainties of both mixtures are reported in Table 2. More detailed description of the filling steps, premixture realizations, and gas chromatographic validations of the gravimetric procedure used for the preparation of the mixtures (according to ISO 6142-1 [26]) can be followed on the Experimental section and Appendix A of our previous work [19].

Remark that the $H_2$-enriched natural gas mixture is not a real blend of the 11 M synthetic mixture (the composition of several components, such as ethane, propane, nitrogen, and carbon dioxide, is rather different). The 11 M mixture is a primary certificate standard mixture proposed by the Physical-Technical Federal Institute (Physikalisch-Technische Bundesanstalt, PTB), whereas the $H_2$-enriched natural gas mixture is a mixture proposed by the Consultative Committee for Amount of Substance (Comité consultative pour la quantité de matière, CCQM) for key comparisons.

**Table 2.** Normalized molar composition $x_i$ and expanded ($k$ = 2) uncertainty $U(x_i)$ of the two natural gas mixtures studied in this work.

| Components | CAS-number | 11 M synthetic natural gas[a] | | $H_2$-enriched natural gas[b] | |
|---|---|---|---|---|---|
| | | $10^2 \cdot x_i$ / mol·mol$^{-1}$ | $10^2 \cdot U(x_i)$ / mol·mol$^{-1}$ | $10^2 \cdot x_i$ / mol·mol$^{-1}$ | $10^2 \cdot U(x_i)$ / mol·mol$^{-1}$ |
| Methane | 74-82-8 | 87.6637 | 0.0035 | 78.8212 | 0.0038 |
| Ethane | 74-84-0 | 4.22521 | 0.00046 | 0.75736 | 0.00016 |
| Propane | 74-98-6 | 1.0490 | 0.0021 | 0.297078 | 0.000089 |
| Butane | 106-97-8 | 0.21265 | 0.00010 | 0.200439 | 0.000098 |
| Isobutane | 75-28-5 | 0.210325 | 0.000084 | 0.197954 | 0.000035 |
| Pentane | 109-66-0 | 0.051829 | 0.000027 | 0.050134 | 0.000021 |
| Isopentane | 78-78-4 | 0.052184 | 0.000027 | 0.049928 | 0.000021 |
| Neopentane | 463-82-1 | | | 0.049615 | 0.000031 |
| Hexane | 110-54-3 | 0.052567 | 0.000024 | 0.050708 | 0.000019 |
| Carbon dioxide | 124-38-9 | 1.62285 | 0.00030 | 4.00108 | 0.00028 |
| Nitrogen | 7727-37-9 | 4.32170 | 0.00078 | 12.01783 | 0.00077 |
| Oxygen | 7782-44-7 | 0.53801 | 0.00011 | | |
| Helium | 7440-59-7 | | | 0.49690 | 0.00030 |
| Hydrogen | 1333-74-0 | | | 3.0097 | 0.0013 |

(a) BAM bottle no./BAM mixture label: C49358-090825/BAM-G420.

(b) BAM bottle no./BAM mixture label: 8099-160905/CCQM-K118.

Note that the reported gravimetric compositions [19] include traces of existing impurities originating from the pure components as well, which are not specifically listed in Table 2 presented in this work. However, their contribution was accounted in the mole fraction uncertainty given in Table 2. In addition, neopentane (2,2-dimethylpropane) is considered neither in the AGA8-DC92 EoS [20,21] nor in the GERG-2008 EoS [22,23]. Hence, it must be treated as a trace component and its composition added to the composition of n-pentane as indicated in ISO 20765-2 [27], whereby the mixture model is still valid should the composition of all of these trace components not exceed a mole percentage of 0.05, as it is the case here.

Finally, the two mixtures were homogenized by heating and rolling at BAM before taking the measurements.

**2.2 Experimental setup.**

The experimental apparatus is the same as described before [28,29], thus a brief description is given here for completeness. The main part is the acoustic resonant cavity made with a spherical shape within tolerances better than 1 μm in 321 austenitic stainless steel. Two hollow hemispheres were welded by electron beam. It was designed to be a pressure-tight shell of 268 $cm^3$ with a wall of a thickness $b$ = 12.5 mm and a nominal internal radius $a$ = 40 mm. Dependence of the latter $a(p,T)$ with pressure and temperature has been determined previously [29] by speed of sound measurements in argon, a gas with an established equation of state [30].

Four ports are opened in the cavity wall. Two ports with 1.5 mm radius in the northern hemisphere form an angle of 45° with the north pole to accommodate the plugs where the acoustic transducers (source and detector) are placed. The other two ports with a radius $r_0$ = 0.8 mm are to provide access to the inlet gas duct of length $L_1$ = 2.3 m and a blind duct of length $L_2$ = 0.035 m, respectively, the latter used for measuring the speed of sound with gas flow but deactivated during this work. They are located in the north and south poles, respectively. This cavity is located inside a vacuum-tight vessel which in turn is submerged in a Dewar vessel filled with ethanol and cooled by a stirred thermal bath at $T$ = –22.5ºC. Three band resistors around the copper block that clamp the north pole of the cavity, the side and the bottom of the shell heat the resonator to the desired experimental temperature which is controlled by a PID (proportional-integral-derivative) loop, with a thermal stability in the order of 1 mK during the entire experimental runs.

A 40 V peak-to-peak signal amplified up to 180 V is sent from the wave generator (3225B function generator, HP) across a set of electrical feed-throughs in the top plate of the vessel to the source capacitance-type acoustic transducer, constructed by a thin polyamide film of 12 µm thickness and gold-plated in the side facing the interior of the acoustic cavity. It generates an acoustic signal up to 20 mPa which is detected by an equal transducer at twice the driven frequency $f$, plugged to a high-input impedance amplifier in order to operate at constant charge. The detected signal is driven by triaxial cables across the feed-throughs to a lock-In amplifier (SR850 DSP Lock-In, SRS) working at differential mode and phase-locked with the wave synthesizer. The signal is then decomposed to the in-phase $u$ and quadrature $v$ components and fitted to a Lorentzian function:

$$u + \mathrm{i}v = \frac{A^*}{\left(F^2 - f^2\right)} + B^* + C^* f \tag{1}$$

where $A^*$ is a term proportional to the amplitude of the acoustic field inside the resonator and the sensitivity of the detector acoustic transducer, $B^*$ and $C^*$ are complex terms that account for the constant and linear background level, and $F = f_{0n} + \mathrm{i}g_{0n}$ stands for the complex resonance frequency with $f_{0n}$ and $g_{0n}$ equal to the experimental resonance frequency and halfwidth of the radial non-degenerate ($l = 0,n$) mode, respectively.

The thermodynamic state is determined by the mean of the temperature readings from two standard platinum reference thermometers (25.5 Ω SPRT 162D, Rosemount) located in the northern and souther hemispheres of the acoustic cavity and monitored by an AC resistance bridge (F18 automatic bridge, ASL). This system has been calibrated in our accredited facilities on the ITS-90 [31] with an estimated expanded ($k$ = 2) uncertainty $U_{\mathrm{cal}}$ ($T$) = 4 mK. The pressure is determined by two piezoelectric quartz transducers connected to the top of the inlet gas duct, which cover the ranges (0 to 2) MPa (Digiquartz 2003A-101-CE, Paroscientific) and (0 to 20) MPa (Digiquartz 43K-101, Paroscientific). They have also been calibrated against a dead-weight pneumatic balance in our laboratory. The estimated expanded ($k$ = 2) uncertainty in pressure is $U_{\mathrm{cal}} = (7.5\cdot10^{-5}\cdot(p/\mathrm{MPa}) + 2\cdot10^{-4})$ MPa.

**2.3 Measurement procedure and acoustic model.**

Several consecutive repetitions of the resonance frequency $f_{0\mathrm{n}}$ and halfwidth $g_{0\mathrm{n}}$ of the first five purely radial acoustic modes (0,2), (0,3), (0,4), (0,5), and (0,6) are recorded at frequencies between (6.65 to 31.31) kHz, starting at the highest pressure of about 13 MPa down to approximately 0.1 MPa, upon reducing the pressure in 1.5-MPa steps and, when a stable temperature condition has been met at every pressure, typically after eight hours. Then, these measurements at each experimental ($p_{\mathrm{i}}$,$T_{\mathrm{i,exp}}$) state are corrected to specify them to the same reference temperature $T_{\mathrm{ref}}$, for every isotherm by multiplying them with the term $w(p_{\mathrm{i}},T_{\mathrm{ref}})/w(p_{\mathrm{i}},T_{\mathrm{i,exp}})$. Finally, from the mean of $f_{0\mathrm{n}}$ and $g_{0\mathrm{n}}$, a speed of sound value $w_{0\mathrm{n}}$ is determined for every mode as shown in Equation (2):

$$w_{0n} = 2\pi a \frac{f_{0n} - \Delta f}{z_{0n}} \tag{2}$$

where $a$ stands for the internal radius of the resonance cavity obtained as discussed above, $z_{0\mathrm{n}}$ stands for the $n$-th zero of the spherical Bessel first derivative of order $l$ = 0, and $\Delta f$ is the term that accounts for the sum of all frequency corrections due to the non-zero acoustic wall admittance and imperfect geometry of the cavity, respectively. The former case comprises the perturbations induced by the heat exchange in the thermal boundary layer [32], the coupling of the motion of the fluid and shell motion [33], and the presence of the source and detector acoustic transducers [24]. The latter case includes the perturbations due to the two gas ducts [34]. Other geometrical imperfections, such as slits in the equatorial joint or around the transducer plugs, are demonstrated to be negligible [32], with corrections less than 1 part in $10^6$, especially for the radial acoustic modes, which are not sensitive to perturbations of the perfect sphericity on first order [35]. Standard models have been used to calculate these corrections, and the specific expressions of every term have been reported elsewhere [36]. REFPROP 10.0 [25], computed the thermodynamic and transport properties of the gas mixtures required by these correction models using the GERG-2008 EoS [22,23]. The elastic and thermal properties of the stainless steel of the cavity wall were obtained from the correlations of [37] and [38,39], respectively, when required. The magnitude of the overall frequency perturbations $\Delta f/f$ that must be subtracted to the experimental frequencies $f_{0n}$ ranges from (−130 to −250) parts in $10^6$ for the (0,2) mode of the $H_2$-enriched mixture at $T$ = 260 K up to (−50 to +700) parts in $10^6$ for the (0,6) mode of the 11 M synthetic mixture at $T$ = 350 K, which is in the order of the estimated expanded ($k$ = 2) uncertainty $U_r(w_{exp})$ ~ 200 parts in $10^6$ for the two mixtures as described below and, hence, not a negligible quantity.

The validity of the applied acoustic model is assessed by the extent of the relative dispersion of the speed of sound $\Delta w/w = (w_{0n} - \langle w \rangle)/\langle w \rangle$ from every mode $w_{0n}$ around their mean $\langle w \rangle$ and from the relative excess halfwidth $\Delta g/f$ for every acoustic mode:

$$\left.\frac{\Delta g}{f}\right|_{0n} = \frac{g_{0n} - \left(g_{th} + g_0 + g_{bulk}\right)}{f_{0n}} \tag{3}$$

where $g_{th}$ is the contribution accounting for energy losses in the thermal boundary layer, $g_0$ is the term accounting for energy losses in the gas tubes, and $g_{bulk}$ is the classical viscothermal dissipation in the bulk of the fluid. Figure 2 depicts $\Delta g/f$ and Figure 3 displays $\Delta w/w$ as a function of pressure for the two studied mixtures at the highest experimental isotherm $T$ = 350 K. As can be seen from Figure 2, the relative excess halfwidths are always in the order of 30 parts in $10^6$ for the modes (0,2), (0,3), and (0,4). By contrast, this magnitude is larger than the experimental expanded ($k$ = 2) uncertainty $U_r(w_{exp})$ ~ 200 parts in $10^6$ for the mode (0,5) at the lowest and highest pressures and for the mode (0,6) at all the pressures investigated.

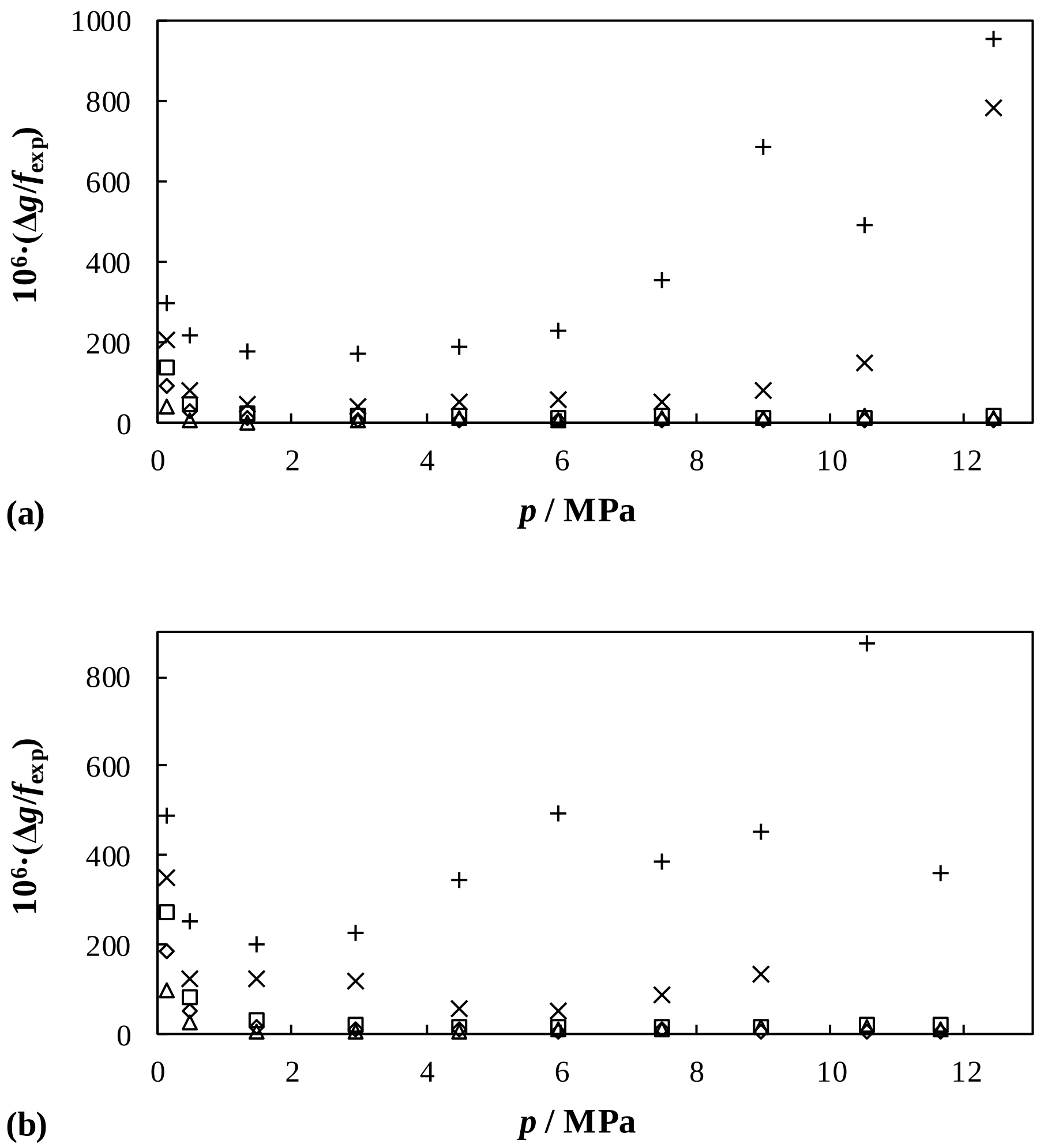

**Figure 2.** Relative excess halfwidths $\Delta g/f$ for: a) the 11 M synthetic natural gas and b) the $H_2$-enriched natural gas mixtures, at $T$ = 350 K and modes △ (0,2), ◇ (0,3), □ (0,4), × (0,5), + (0,6).

In addition, Figure 3 shows that modes (0,5) and (0,6) clearly disagree from the values of the other modes at several pressures, with speed of sound results outside the standard deviation of the mean of 15 parts in $10^6$ for the modes (0,2), (0,3), and (0,4). Both reasons indicate that there are unknown imperfections of the used acoustic model, hence the model cannot perfectly reproduce the experimental situation inside the acoustic cavity. Similar results are found for the remaining isotherms. Thus, mode (0,6) at all the isotherms investigated, as well as mode (0,5) at (300, 325, and 350) K, have been skipped from the following calculations for the two studied mixtures.

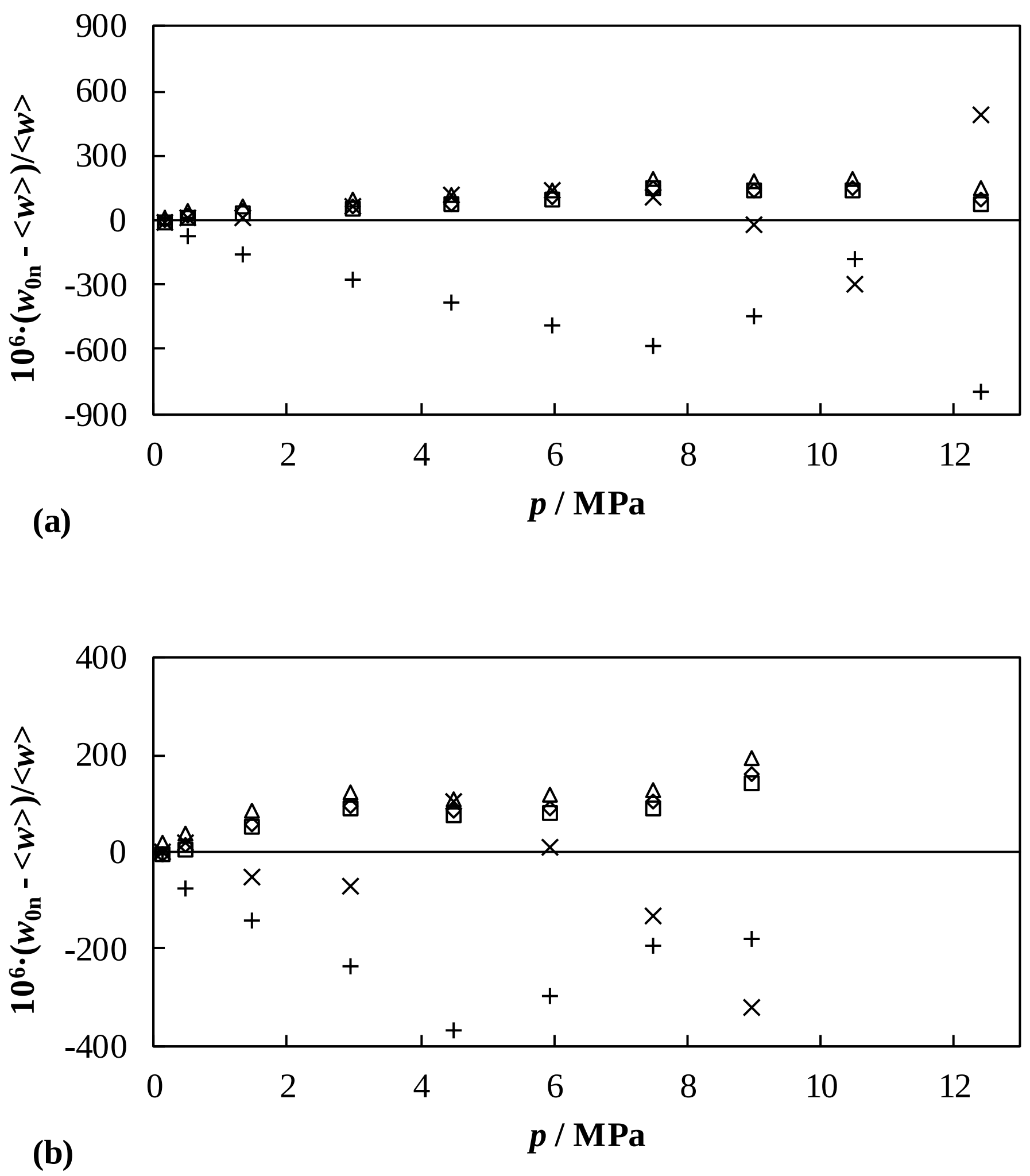

**Figure 3.** Relative dispersion of the speed of sound $\Delta w/w = (w_{0n} - \langle w \rangle)/\langle w \rangle$, where $\langle w \rangle$ is the mean value from modes (0,2) to (0,6), as a function of pressure for: a) the 11 M synthetic natural gas and b) the $H_2$-enriched natural gas mixtures, at $T$ = 350 K and modes △ (0,2), ◇ (0,3), □ (0,4), × (0,5), + (0,6).

**3. Results and uncertainty.**

One hundred experimental $w_{exp}(p,T,\bar{x})$ points have been measured at pressures $p$ from (0.1 up to 13) MPa and temperatures $T$ = (260, 273.16, 300, 325, and 350) K. Tables 3 and 4 show the data sets for the 11 M synthetic natural gas and the $H_2$-enriched natural gas, respectively. Each point reports the mean of the speed of sounds determined from Equation (2) using the remaining radial $(0,n)$ resonance modes after rejection of those modes whose relative excess halfwidths $\Delta g_{0n}/f_{0n}$ are significantly larger and whose relative speed of sound dispersion $\Delta w/w$ clearly differs from the others, as discussed before.

**Table 3.** Speeds of sound $w_{exp}(p,T)$ for the 11 M synthetic natural gas mixture with their relative expanded ($k$ = 2) uncertainty[a] and relative deviations $(w_{exp} - w_{EoS})/w_{EoS} = \Delta w/w_{EoS}$ from the speed of sound predicted by the AGA8-DC92 EoS [20,21] and the GERG-2008 EoS [22,23].

| $p$ / MPa | $w_{exp}$ / m·s$^{-1}$ | $10^6·(w_{exp} - w_{AGA})/w_{AGA}$ | $10^6·(w_{exp} - w_{GERG})/w_{GERG}$ | $p$ / MPa | $w_{exp}$ / m·s$^{-1}$ | $10^6·(w_{exp} - w_{AGA})/w_{AGA}$ | $10^6·(w_{exp} - w_{GERG})/w_{GERG}$ |
|---|---|---|---|---|---|---|---|
| | | $T$ = 260.00 K | | | | $T$ = 325.00 K | |
| 0.13876 | 392.127 | −365 | −418 | 0.12678 | 434.631 | −131 | −157 |
| 0.47800 | 390.049 | −243 | −328 | 0.46853 | 433.835 | 7 | −59 |
| 1.47237 | 384.027 | −122 | −288 | 1.44685 | 431.680 | 192 | 20 |
| 2.98914 | 375.498 | −40 | −289 | 2.91962 | 429.075 | 394 | 68 |
| 4.54980 | 368.152 | −60 | −278 | 4.50284 | 427.322 | 590 | 112 |
| 6.07522 | 363.313 | −204 | −235 | 6.02004 | 426.847 | 776 | 166 |
| 7.63878 | 362.167 | −375 | −141 | 7.46732 | 427.656 | 950 | 227 |
| 9.18078 | 366.524 | −477 | −121 | 9.04454 | 430.111 | 1138 | 302 |
| 10.62418 | 376.894 | −520 | −341 | 10.53238 | 434.066 | 1315 | 383 |
| 11.89247 | 391.359 | −559 | −802 | 12.08749 | 439.989 | 1479 | 464 |
| | | $T$ = 273.16 K | | | | $T$ = 350.00 K | |
| 0.12548 | 401.437 | −314 | −353 | 0.14448 | 449.138 | −85 | −114 |
| 0.48846 | 399.604 | −130 | −210 | 0.49294 | 448.617 | 25 | −47 |
| 1.48187 | 394.644 | 13 | −174 | 1.31844 | 447.482 | 168 | 3 |
| 3.00368 | 387.794 | 178 | −155 | 2.97716 | 445.865 | 388 | 56 |
| 4.47096 | 382.451 | 290 | −134 | 4.47555 | 445.272 | 551 | 80 |
| 5.91417 | 378.958 | 362 | −79 | 5.97484 | 445.613 | 703 | 101 |
| 7.43740 | 377.911 | 416 | 3 | 7.50472 | 447.030 | 864 | 142 |
| 9.03526 | 380.706 | 534 | 116 | 9.01878 | 449.576 | 1026 | 193 |
| 10.62452 | 388.354 | 675 | 149 | 10.50800 | 453.256 | 1184 | 257 |
| 12.57160 | 405.012 | 756 | −17 | 12.41850 | 459.739 | 1361 | 342 |
| | | $T$ = 300.00 K | | | | | |

| | | | |
|---|---|---|---|
| 0.13664 | 419.154 | −217 | −250 |
| 0.47079 | 418.020 | −31 | −102 |
| 1.44929 | 414.762 | 147 | −33 |
| 2.98746 | 410.383 | 339 | −9 |
| 4.46523 | 407.305 | 518 | 26 |
| 6.01409 | 405.583 | 693 | 78 |
| 8.02558 | 406.238 | 926 | 183 |
| 9.07127 | 408.105 | 1050 | 240 |
| 10.54125 | 412.733 | 1240 | 329 |
| 11.44834 | 416.825 | 1350 | 382 |

[a] Expanded uncertainties ($k$ = 2): $U(p) = (7.5\cdot10^{-5}$ ($p$/Pa) + 200) Pa; $U(T)$ = 4 mK; $U_r(w)$ = $1.9\cdot10^{-4}$ m·s$^{-1}$/ m·s$^{-1}$.

**Table 4.** Speeds of sound $w_{exp}(p,T)$ for the $H_2$-enriched natural gas mixture with their relative expanded ($k$ = 2) uncertainty[a] and relative deviations $(w_{exp} - w_{EoS})/w_{EoS}$ from the speed of sound predicted by the AGA8-DC92 EoS [20,21] and the GERG-2008 EoS [22,23].

| $p$ / MPa | $w_{exp}$ / m·s$^{-1}$ | $10^6\cdot(w_{exp} - w_{AGA})/w_{AGA}$ | $10^6\cdot(w_{exp} - w_{GERG})/w_{GERG}$ | $p$ / MPa | $w_{exp}$ / m·s$^{-1}$ | $10^6\cdot(w_{exp} - w_{AGA})/w_{AGA}$ | $10^6\cdot(w_{exp} - w_{GERG})/w_{GERG}$ |
|---|---|---|---|---|---|---|---|
| $T$ = 260.00 K | | | | $T$ = 325.00 K | | | |
| 0.14773 | 390.998 | −562 | −597 | 0.12437 | 433.728 | −325 | −336 |
| 0.47021 | 389.662 | −295 | −356 | 0.49608 | 433.303 | −118 | −161 |
| 1.47966 | 385.458 | −101 | −224 | 1.42737 | 432.303 | 104 | −16 |
| 2.98774 | 379.964 | 87 | −88 | 2.97052 | 431.250 | 329 | 107 |
| 4.37248 | 376.083 | 176 | 14 | 4.47759 | 431.088 | 494 | 207 |
| 6.00833 | 373.541 | 284 | 201 | 5.92270 | 431.850 | 658 | 331 |
| 7.61769 | 373.929 | 425 | 377 | 7.47048 | 433.756 | 840 | 484 |
| 9.13792 | 377.666 | 694 | 454 | 9.04792 | 436.950 | 1042 | 654 |
| 10.67254 | 385.328 | 1050 | 290 | 10.56114 | 441.256 | 1234 | 794 |
| 12.31649 | 398.132 | 1368 | −237 | 11.92213 | 446.191 | 1389 | 881 |
| | | $T$ = 273.16 K | | | | $T$ = 350.00 K | |
| 0.13451 | 400.369 | −355 | −409 | 0.13352 | 448.308 | −372 | −384 |
| 0.49849 | 399.129 | −150 | −235 | 0.47718 | 448.173 | −145 | −192 |
| 1.47309 | 395.917 | 91 | −67 | 1.47412 | 447.782 | 80 | −57 |
| 2.97290 | 391.704 | 289 | 46 | 2.95101 | 447.699 | 270 | 30 |
| 4.50887 | 388.664 | 450 | 167 | 4.48238 | 448.348 | 424 | 109 |
| 6.07316 | 387.302 | 612 | 326 | 5.96072 | 449.752 | 565 | 207 |
| 7.50730 | 387.979 | 794 | 492 | 7.50214 | 452.081 | 713 | 330 |
| 9.12321 | 391.385 | 1061 | 626 | 8.98999 | 455.210 | 868 | 467 |
| 10.58929 | 397.216 | 1344 | 621 | 10.54386 | 459.422 | 1026 | 601 |
| 12.12160 | 406.278 | 1615 | 447 | 11.64110 | 462.987 | 1135 | 688 |
| | | $T$ = 300.00 K | | | | | |
| 0.14344 | 418.185 | −330 | −349 | | | | |
| 0.47956 | 417.502 | −113 | −160 | | | | |
| 1.58308 | 415.333 | 111 | −21 | | | | |
| 2.97871 | 413.240 | 303 | 85 | | | | |
| 4.49573 | 411.973 | 481 | 200 | | | | |
| 6.01298 | 411.946 | 653 | 336 | | | | |
| 7.51230 | 413.322 | 833 | 484 | | | | |
| 9.02840 | 416.308 | 1053 | 647 | | | | |
| 10.58955 | 421.186 | 1296 | 772 | | | | |
| 12.14970 | 427.961 | 1516 | 815 | | | | |

[a] Expanded uncertainties ($k$ = 2): $U(p)$ = (7.5·10$^{-5}$ ($p$/Pa) + 200) Pa; $U(T)$ = 4 mK; $U_r(w)$ = 1.8·10$^{-4}$ m·s$^{-1}$/ m·s$^{-1}$.

Tables 5 and 6 report the specified uncertainty contributions to the speed of sound of the 11 M synthetic natural gas mixture and the $H_2$-enriched natural gas mixture, respectively, as the average of the uncertainties of all speed of sound datasets. The square root of the sum of the squares of these components yields the overall relative expanded ($k$ = 2) uncertainty of the speed of sound $U_r(w_{exp})$ of 190 parts in $10^6$ (0.019 %) for the 11 M synthetic natural gas mixture and 180 parts in $10^6$ (0.018 %) for the $H_2$-enriched natural gas mixture, respectively. As predicted, the most significant contribution to $U_r(w_{exp})$ is due to the geometrical characterization of the resonance cavity by means of speed of sound measurements in argon to determine the behavior of the internal cavity radius as function of the pressure and temperature $U(a)$, which amounts up to 170 parts in $10^6$. Next is the contribution of the uncertainty of the composition of the gas mixtures $U(x_i)$ given in Table 2, which is lower than (70 and 50) parts in $10^6$ for the 11 M synthetic mixture and the $H_2$-enriched mixture, respectively. Minor contributing terms are: (a) the adequateness of the acoustic model described above and quantified in the order of 10 parts in $10^6$ as the standard deviation of the speed of sound from the different non-rejected modes $U(<w>)$; (b) the imperfect determination of the thermodynamic state, whose contribution is evaluated from the pressure uncertainty $U(p)$ as 3 part in $10^6$ and from the temperature uncertainty $U(T)$ as 5 parts in $10^6$; and (c) the error associated with the fitting of the recorded in-phase and quadrature acoustic signals to the Lorentzian shape function of Equation (1) $U(f_{0n})$, always below 1 part in $10^6$. Detailed descriptions to calculate every part of the uncertainty have been given elsewhere [40,41].

**Table 5.** Uncertainty budget for the speed of sound measurements $w_{exp}$ for the 11 M synthetic natural gas mixture. Unless otherwise specified, uncertainties are indicated with a coverage factor $k$ = 1.

| Source | Magnitude | | Contribution to the speed of sound uncertainty, $10^6 \cdot u_r(w_{exp})$ |
|---|---|---|---|
| Temperature | Calibration | 0.002 K | |
| | Resolution | $7.2 \cdot 10^{-7}$ K | |
| | Repeatability | $5.9 \cdot 10^{-5}$ K | |
| | Gradient (across hemispheres) | $1.3 \cdot 10^{-3}$ K | |
| | Quadrature Sum | $2.5 \cdot 10^{-3}$ K | 4.9 |
| Pressure | Calibration | $(3.75 \cdot 10^{-5} \cdot p + 1 \cdot 10^{-4})$ MPa | |
| | Resolution | $2.9 \cdot 10^{-5}$ MPa | |
| | Repeatability | $1.2 \cdot 10^{-5}$ MPa | |
| | Quadrature Sum | $(1.1 \text{ to } 5.5) \cdot 10^{-4}$ MPa | 2.8 |
| Gas composition | Purity | $9.7 \cdot 10^{-7}$ kg/mol | |
| | Molar mass | $9.9 \cdot 10^{-7}$ kg/mol | |
| | Quadrature Sum | $1.4 \cdot 10^{-6}$ kg/mol | 34 |
| Radius from speed of sound in Ar | Temperature | $1.5 \cdot 10^{-9}$ m | |
| | Pressure | $1.6 \cdot 10^{-10}$ m | |
| | Gas Composition | $4.1 \cdot 10^{-9}$ m | |
| | Frequency fitting | $4.9 \cdot 10^{-7}$ m | |
| | Regression | $1.7 \cdot 10^{-6}$ m | |
| | Equation of State | $2.3 \cdot 10^{-6}$ m | |
| | Dispersion of modes | $2.9 \cdot 10^{-6}$ m | |
| | Quadrature Sum | $4.2 \cdot 10^{-6}$ m | 88 |
| Frequency fitting | | $5.6 \cdot 10^{-3}$ Hz | 0.51 |
| Dispersion of modes | | $5.7 \cdot 10^{-3}$ m·s$^{-1}$ | 14 |
| Quadrature Sum of all contributions to $w_{exp}$ | | | 95 |
| $10^6 \cdot U_r(w_{exp})$ ($k = 2$) | | | 190 |

**Table 6.** Uncertainty budget for the speed of sound measurements $w_{exp}$ for the $H_2$-enriched natural gas mixture. Unless otherwise specified, uncertainties are indicated with a coverage factor $k = 1$.

| Source | Magnitude | | Contribution to the speed of sound uncertainty, $10^6 \cdot u_r(w_{exp})$ |
|---|---|---|---|
| Temperature | Calibration | 0.002 K | |
| | Resolution | $7.2 \cdot 10^{-7}$ K | |
| | Repeatability | $3.5 \cdot 10^{-5}$ K | |
| | Gradient (across hemispheres) | $1.0 \cdot 10^{-3}$ K | |
| | Quadrature Sum | $2.3 \cdot 10^{-3}$ K | 4.4 |
| Pressure | Calibration | $(3.75 \cdot 10^{-5} \cdot p + 1 \cdot 10^{-4})$ MPa | |
| | Resolution | $2.9 \cdot 10^{-5}$ MPa | |
| | Repeatability | $7.9 \cdot 10^{-6}$ MPa | |
| | Quadrature Sum | $(1.1 \text{ to } 5.5) \cdot 10^{-4}$ MPa | 2.1 |
| Gas composition | Purity | $8.7 \cdot 10^{-7}$ kg/mol | |
| | Molar mass | $3.3 \cdot 10^{-7}$ kg/mol | |
| | Quadrature Sum | $9.3 \cdot 10^{-7}$ kg/mol | 23 |
| Radius from speed of sound in Ar | Temperature | $1.5 \cdot 10^{-9}$ m | |
| | Pressure | $1.6 \cdot 10^{-10}$ m | |
| | Gas Composition | $4.1 \cdot 10^{-9}$ m | |
| | Frequency fitting | $4.9 \cdot 10^{-7}$ m | |
| | Regression | $1.7 \cdot 10^{-6}$ m | |
| | Equation of State | $2.3 \cdot 10^{-6}$ m | |
| | Dispersion of modes | $2.9 \cdot 10^{-6}$ m | |
| | Quadrature Sum | $4.2 \cdot 10^{-6}$ m | 88 |
| Frequency fitting | | $7.6 \cdot 10^{-3}$ Hz | 0.69 |
| Dispersion of modes | | $5.3 \cdot 10^{-3}$ m·s$^{-1}$ | 13 |
| Quadrature Sum of all contributions to $w_{exp}$ | | | 92 |
| $10^6 \cdot U_r(w_{exp})$ ($k$ = 2) | | | 180 |

Squared speed of sound data $w^2(p,T)$ have been fitted to the standard form of the acoustic virial equation as an expansion series of pressure $p$ at each experimental temperature $T$ as:

$$w^2(p,T) = A_0(T) + A_1(T)\,p + A_2(T)\,p^2 + A_3(T)\,p^3 + A_4(T)\,p^4 + A_5(T)\,p^5 \qquad (4)$$

providing that: (i) the residuals of the regression are randomly distributed within the above specified expanded ($k$ = 2) uncertainty of the speed of sound, as it is shown in Figure 4; and (ii) the fitting parameters $A_i$, which are given in Table 7, are significant, i.e., their expanded ($k$ = 2) uncertainties evaluated following the Monte Carlo method [42] are at least one order of magnitude lower than the parameters themselves.

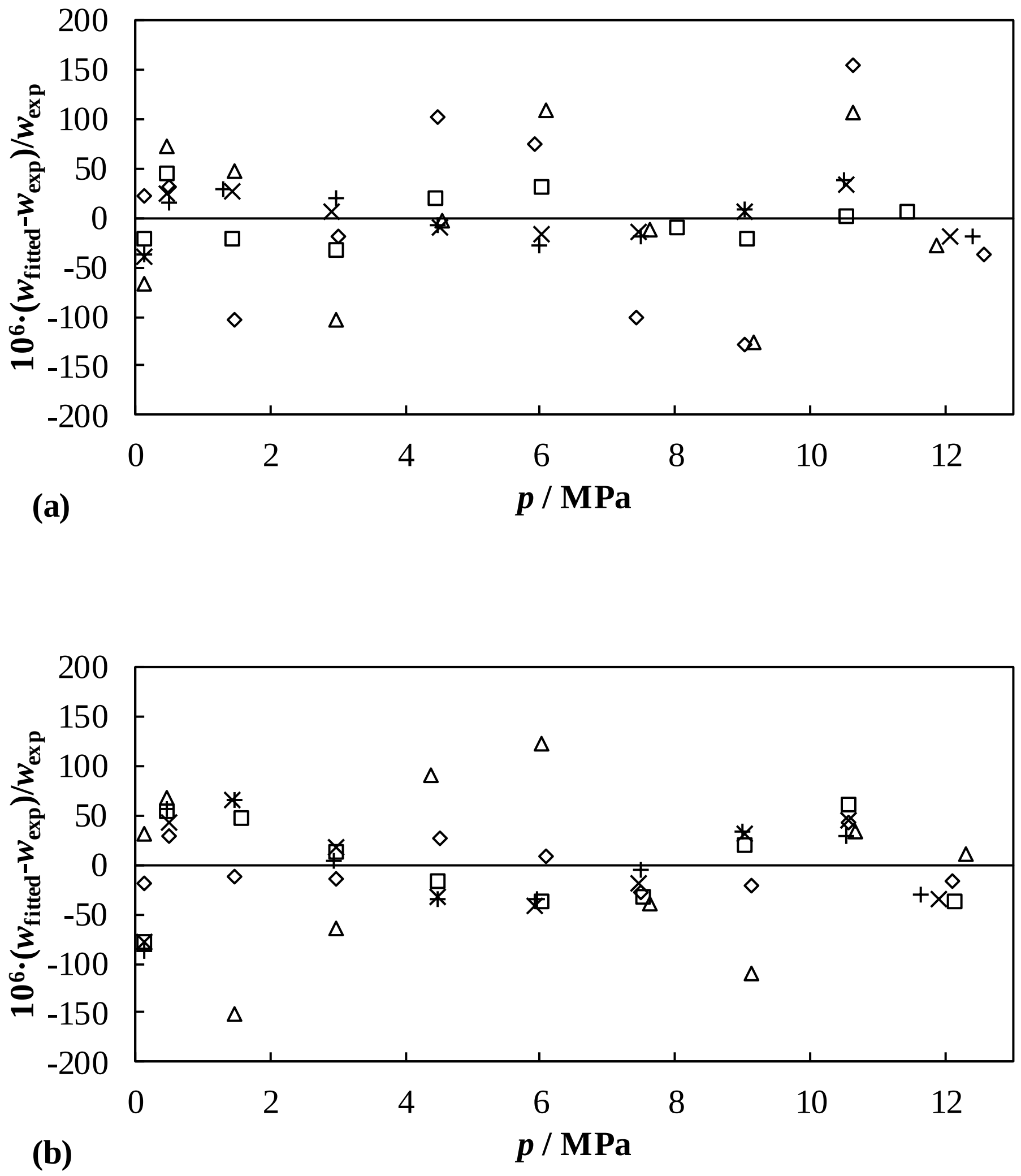

**Figure 4.** Residual plots $\Delta w = (w_{\mathrm{fitted}} - w_{\mathrm{exp}})/w_{\mathrm{exp}}$ from the values regressed to Eq. (4) as a function of the pressure for: a) the 11 M synthetic natural gas and b) the $H_2$-enriched natural gas mixtures, at temperatures $T$ = △ 260 K, ◇ 273.16 K, □ 300 K, × 325 K, + 350 K.

Then, it is concluded that a polynomial of third order is required for all the isotherms except from the isotherms at $T$ = (260 and 273.16) K for the 11M synthetic mixture, where a polynomial of fifth and a fourth order, respectively is needed. This approach results in relative residuals not higher than 150 parts in $10^6$ in any case, with root mean squares below 90 parts in $10^6$ for all the temperatures, which are within $U_r(w_{\mathrm{exp}})$ = (190 and 180) parts in $10^6$ for the 11 M synthetic and the $H_2$-enriched mixtures, respectively. We must remark that the estimated expanded ($k$ = 2) uncertainties of $w^2(p,T)$ were used as weights in the fitting to Equation (4).

**Table 7.** Fitting parameters $A_i(T)$ of the squared speed of sound to Eq. (4), their corresponding expanded ($k$ = 2) uncertainties determined by the Monte Carlo method and the root mean square ($\Delta_{RMS}$) of the residuals of the fitting.

| $T$ / K | $A_0(T)$ / m$^2$·s$^{-2}$ | $10^5$·$A_1(T)$ / m$^2$·s$^{-2}$·Pa$^{-1}$ | $10^{12}$·$A_2(T)$ / m$^2$·s$^{-2}$·Pa$^{-2}$ | $10^{19}$·$A_3(T)$ / m$^2$·s$^{-2}$·Pa$^{-3}$ | $10^{26}$·$A_4(T)$ / m$^2$·s$^{-2}$·Pa$^{-4}$ | $10^{32}$·$A_5(T)$ / m$^2$·s$^{-2}$·Pa$^{-5}$ | $\Delta_{RMS}$ of residuals / ppm |
|---|---|---|---|---|---|---|---|
| | | | | 11 M synthetic natural gas | | | |
| 260.00 | 154482 ± 16 | −504.9 ± 3.6 | 216.0 ± 20.0 | −249.0 ± 44.0 | 499.0 ± 41.0 | −12.5 ± 1.3 | 79 |
| 273.16 | 161656 ± 17 | −409.5 ± 2.5 | 58.6 ± 8.6 | 175.0 ± 11.0 | 39.5 ± 4.1 | | 90 |
| 300.00 | 176102 ± 17 | −297.4 ± 1.6 | 97.8 ± 3.4 | 125.7 ± 1.9 | | | 24 |
| 325.00 | 189192 ± 20 | −216.9 ± 1.7 | 125.8 ± 3.6 | 69.4 ± 1.9 | | | 22 |
| 350.00 | 201951 ± 21 | −148.4 ± 1.8 | 129.4 ± 3.6 | 41.2 ± 1.9 | | | 24 |
| | | | | $H_2$-enriched natural gas | | | |
| 260.00 | 153355 ± 11 | −329.70 ± 0.94 | 34.7 ± 1.9 | 216.7 ± 1.0 | | | 84 |
| 273.16 | 160679 ± 15 | −282.6 ± 1.3 | 88.1 ± 2.6 | 144.2 ± 1.4 | | | 24 |
| 300.00 | 175179 ± 17 | −191.6 ± 1.4 | 124.0 ± 2.9 | 72.3 ± 1.6 | | | 45 |
| 325.00 | 188296 ± 19 | −119.2 ± 1.7 | 124.9 ± 3.6 | 42.8 ± 2.0 | | | 45 |
| 350.00 | 201091 ± 21 | −58.9 ± 1.9 | 115.9 ± 4.1 | 28.1 ± 2.3 | | | 45 |

In addition, it can be demonstrated that [40]: (a) from $A_0$ the heat capacity ratio $\gamma^{pg}$ as perfect-gas (zero pressure conditions) is obtained as $\gamma^{pg} = M \cdot A_0/(R \cdot T)$, whereby the molar isochoric heat capacity $C_{V,m}^{pg}$ as perfect-gas is $C_{V,m}^{pg} = R/(\gamma^{pg} - 1)$, and the molar isobaric heat capacity $C_{p,m}^{pg}$ as perfect-gas is $C_{p,m}^{pg} = \gamma^{pg} \cdot C_{V,m}^{pg}$, where $M$ stands for the molar mass of the mixture and $R$ for the molar gas constant; (b) from $A_1$ the second acoustic virial coefficient $\beta_a$ is derived as $\beta_a = R \cdot T \cdot A_1/A_0$; and (c) from $A_2$ the third acoustic virial coefficient $\gamma_a$ is determined as $\gamma_a = (R \cdot T)^2 \cdot A_2/A_0 + \beta_a \cdot B(T)$, where $B(T)$ stands for the second density virial coefficient. Table 8 shows the results of $\gamma^{pg}$, $C_{p,m}^{pg}$, $\beta_a$, and $\gamma_a$ obtained from these regressions together with their relative expanded ($k$ = 2) uncertainties, which includes the contributions from the uncertainty of temperature, molar mass, and fitted $A_i$ parameters (note that after the redefinition of the kelvin, the uncertainty of $R$ is zero [43,44]).

**Table 8.** Adiabatic coefficient $\gamma^{pg}$, isobaric heat capacity $C_{p,m}^{pg}$, acoustic second virial coefficient $\beta_a$, and acoustic third virial coefficient $\gamma_a$ for the two natural gas mixtures analyzed in this work, with their corresponding relative expanded ($k$ = 2) uncertainty, and comparison with AGA8-DC92 EoS [20,21] and GERG-2008 EoS [22,23]. The superscript *pg* indicates perfect-gas property.

| $T$ / K | $\gamma^{pg}$ | $10^2 \cdot U_{exp}(\gamma^{pg})$ | $10^2 \cdot \Delta\gamma^{pg}{}_{AGA}{}^{(*)}$ | $10^2 \cdot \Delta\gamma^{pg}{}_{GERG}{}^{(*)}$ | $C_{p,m}^{pg}$ / J·mol$^{-1}$·K$^{-1}$ | $10^2 \cdot U_{exp}(C_{p,m}^{pg})$ | $10^2 \cdot \Delta C_{p,m\,AGA}^{pg\,(*)}$ | $10^2 \cdot \Delta C_{p,m\,GERG}^{pg\,(*)}$ |
|---|---|---|---|---|---|---|---|---|
| | | | | 11 M synthetic natural gas | | | | |
| 260.00 | 1.30488 | 0.018 | −0.056 | −0.060 | 35.586 | 0.081 | 0.19 | 0.20 |
| 273.16 | 1.29970 | 0.018 | −0.083 | −0.086 | 36.057 | 0.083 | 0.28 | 0.29 |
| 300.00 | 1.28917 | 0.019 | −0.052 | −0.053 | 37.068 | 0.083 | 0.18 | 0.18 |
| 325.00 | 1.27845 | 0.019 | −0.025 | −0.026 | 38.174 | 0.087 | 0.092 | 0.093 |
| 350.00 | 1.26720 | 0.019 | −0.016 | −0.016 | 39.432 | 0.089 | 0.059 | 0.061 |
| | | | | $H_2$-enriched natural gas | | | | |
| 260.00 | 1.31914 | 0.012 | −0.14 | −0.15 | 34.367 | 0.053 | 0.45 | 0.46 |
| 273.16 | 1.31555 | 0.014 | −0.087 | −0.089 | 34.663 | 0.059 | 0.27 | 0.28 |
| 300.00 | 1.30595 | 0.014 | −0.059 | −0.059 | 35.491 | 0.061 | 0.19 | 0.19 |
| 325.00 | 1.29575 | 0.014 | −0.057 | −0.055 | 36.427 | 0.065 | 0.19 | 0.18 |
| 350.00 | 1.28496 | 0.014 | −0.065 | −0.063 | 37.492 | 0.067 | 0.23 | 0.22 |

| | $10^7\cdot\beta_a$ / $m^3\cdot mol^{-1}$ | $10^2\cdot U_{exp}(\beta_a)$ | $10^2\cdot\Delta\beta_{a,AGA}$(*) | $10^2\cdot\Delta\beta_{a,GERG}$(*) | $10^{10}\cdot\gamma_a$ / $(m^3\cdot mol^{-1})^2$ | $10^2\cdot U_{exp}(\gamma_a)$ | $10^2\cdot\Delta\gamma_{a,AGA}$(*) | $10^2\cdot\Delta\gamma_{a,GERG}$(*) |
|---|---|---|---|---|---|---|---|---|
| | | | | 11 M synthetic natural gas | | | | |
| 260.00 | −706.5 | 0.71 | 1.6 | 2.2 | 113.4 | 5.3 | 42 | 45 |
| 273.16 | −575.3 | 0.59 | −4.4 | −3.6 | 53.5 | 5.2 | −30 | −29 |
| 300.00 | −421.3 | 0.52 | −4.7 | −3.5 | 54.8 | 2.2 | −22 | −21 |
| 325.00 | −309.8 | 0.77 | −4.0 | −2.0 | 60.5 | 2.3 | −8.7 | −6.8 |
| 350.00 | −213.9 | 1.2 | −4.5 | −1.3 | 60.9 | 2.5 | −4.1 | −1.1 |
| | | | | $H_2$-enriched natural gas | | | | |
| 260.00 | −464.8 | 0.28 | −7.6 | −6.9 | 35.85 | 1.6 | −47 | −45 |
| 273.16 | −399.4 | 0.45 | −5.5 | −4.6 | 47.52 | 1.8 | −26 | −25 |
| 300.00 | −272.8 | 0.73 | −4.8 | −3.3 | 54.2 | 1.9 | −9.8 | −8.3 |
| 325.00 | −171.0 | 1.4 | −7.3 | −4.7 | 53.5 | 2.6 | −6.9 | −5.1 |
| 350.00 | −85.3 | 3.3 | −15 | −9.7 | 50.7 | 3.4 | −8.5 | −6.2 |

(*) $\Delta X_{EoS} = (X_{exp} - X_{EoS})/X_{EoS}$ with $X = \gamma^{pg}$, $C_{p,m}^{pg}$, $\beta_a$, $\gamma_a$; and EoS = AGA8-DC92 [20,21], GERG-2008 [22,23].

## 4. Discussion.

### 4.1 Speed of sound.

The relative deviations of the experimental speeds of sound $w_{exp}$ determined in this research from those evaluated by the reference thermodynamic models used in the industry, AGA8-DC92 EoS [20,21] and GERG-2008 EoS [22,23], are given in Table 3 and 4 and depicted in Figures 5 and 6 for the 11 M synthetic natural gas and the $H_2$-enriched natural gas mixtures, respectively.

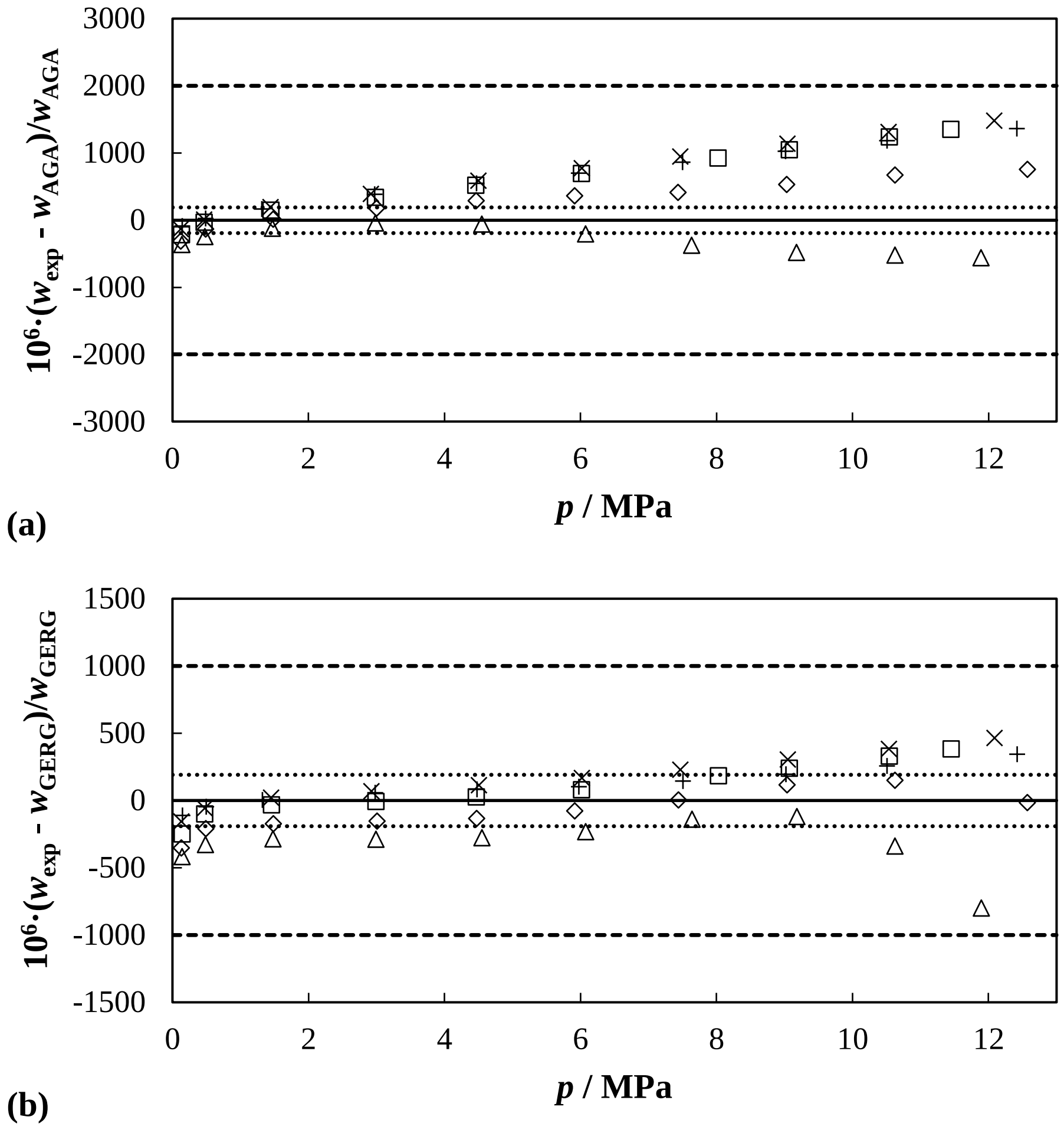

**Figure 5.** Relative deviations of experimental speed of sound $\Delta w = (w_{exp} - w_{EoS})/w_{EoS}$ from speed of sound values calculated from the reference models: a) AGA8-DC92 EoS [20,21] and b) GERG-2008 EoS [22,23] as function of pressure for the 11 M synthetic natural gas mixture at temperatures $T$ = △ 260 K, ◇ 273.16 K, □ 300 K, × 325 K, + 350 K. The dotted lines depict the expanded ($k$ = 2) uncertainty of the experimental speed of sound. The dashed lines represent the uncertainty of the reference models.

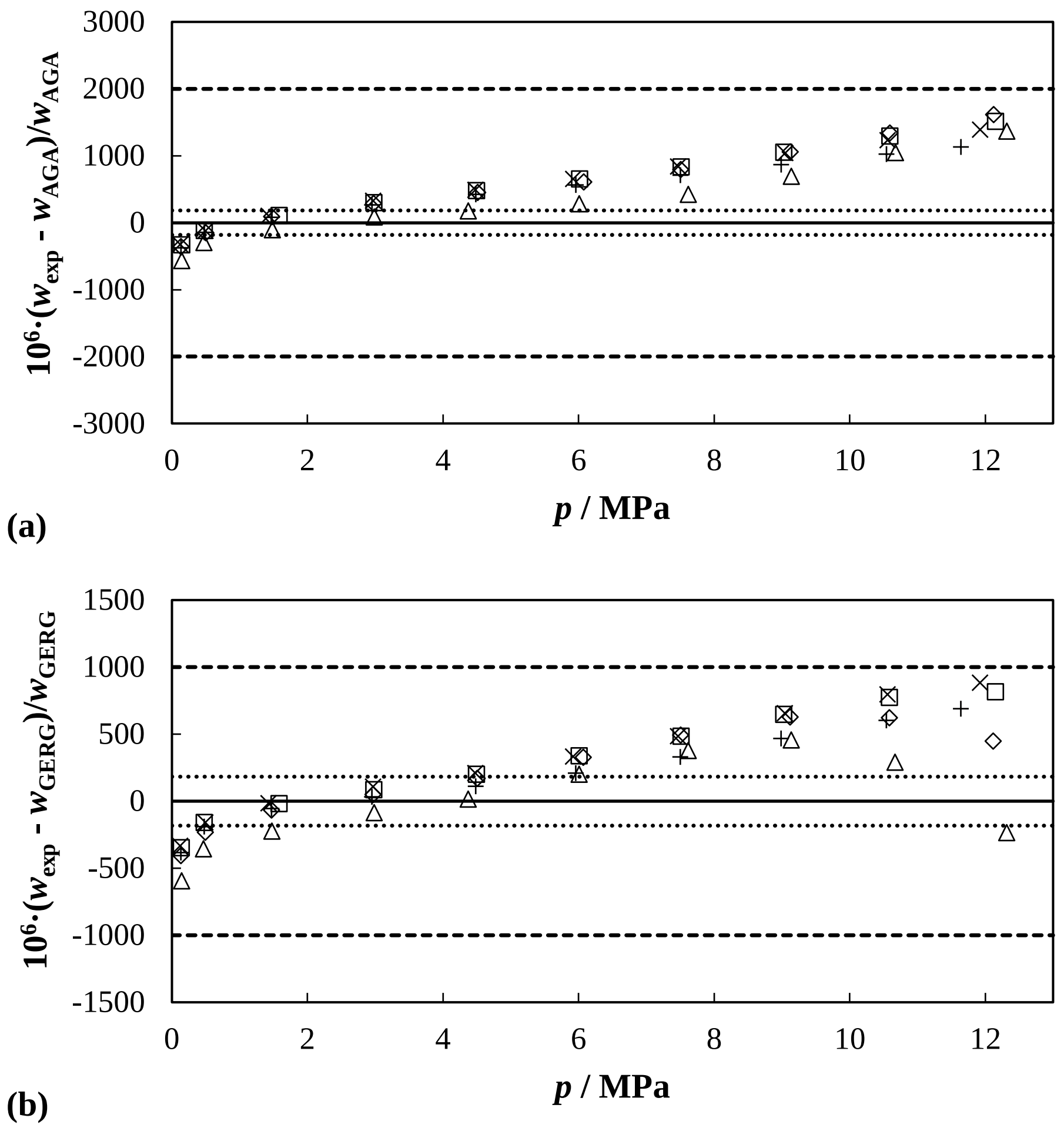

**Figure 6.** Relative deviations of experimental speed of sound $\Delta w = (w_{exp} - w_{EoS})/w_{EoS}$ from speed of sound values calculated from the reference models: a) AGA8-DC92 EoS [20,21] and b) GERG-2008 EoS [22,23] as function of pressure for the $H_2$-enriched natural gas mixture at temperatures $T$ = △ 260 K, ◇ 273 K, □ 300 K, × 325 K, + 350 K. The dotted lines depict the expanded ($k$ = 2) uncertainty of the experimental speed of sound. The dashed lines represent the uncertainty of the reference models.

At all experimental conditions, the deviations of the speed of sound remain within the limit stated by the expanded ($k$ = 2) uncertainty in the speed of sound of both models for natural gas-like mixtures, namely $U_{AGA8\text{-}DC92}(w)$ = 2000 parts in $10^6$ (0.2 %) and $U_{GERG\text{-}2008}(w)$ = 1000 parts in $10^6$ (0.1 %). However, the differences are only explained within the experimental expanded ($k$ = 2) uncertainty $U_r(w_{exp})$ for pressures below (4 and 9) MPa with regards to the AGA and GERG equations, respectively, for the 11 M mixture and below (3 and 5) MPa with respect to the AGA and GERG models, respectively, for the $H_2$-enriched mixture.

The comparison between both mixtures reveals that adding a nominal hydrogen mole percentage of 3 % results in increasing deviations towards higher pressures, with a nearly linear trend which is fairly independent of temperature. This effect is particularly noticeable for the GERG-2008 EoS [22,23] at all the studied isotherms and for the AGA8-DC92 EoS [20,21] at the lowest temperature of 260 K.

Table 9 lists overall indicators of the predicting capability of the two models: the average absolute deviations $\Delta_{AAD}$ for the two mixtures, together with the root mean square $\Delta_{RMS}$, the bias $\Delta_{Bias}$, and the maximum deviation $\Delta_{Max}$, which also serve for further comparison with other studies on speed of sound.

**Table 9.** Statistical analysis of the speed of sound data with respect to AGA8-DC92 EoS [20,21] and GERG-2008 EoS [22,23] for the two natural gas mixtures studied in this work. $\Delta_{AAD}$ = average absolute relative deviation, $\Delta_{Bias}$ = average relative deviation, $\Delta_{RMS}$ = root mean square relative deviation, $\Delta_{Max}$ = maximum relative deviation.

| | $10^2$·(Experimental vs AGA) | | | | $10^2$·(Experimental vs GERG) | | | |
|---|---|---|---|---|---|---|---|---|
| | $\Delta_{AAD}$ | $\Delta_{Bias}$ | $\Delta_{RMS}$ | $\Delta_{Max}$ | $\Delta_{AAD}$ | $\Delta_{Bias}$ | $\Delta_{RMS}$ | $\Delta_{Max}$ |
| 11 M synthetic natural gas | 0.053 | 0.037 | 0.064 | 0.11 | 0.019 | −0.0014 | 0.023 | 0.047 |
| $H_2$-enriched natural gas | 0.061 | 0.050 | 0.075 | 0.14 | 0.034 | 0.019 | 0.041 | 0.072 |

As it can be seen, the discrepancies are twice higher for the AGA8-DC92 EoS [20,21] than the GERG-2008 EoS [22,23] with reference to the two mixtures at the same states, although the impact of adding hydrogen is twice more pronounced for the GERG-2008 EoS [22,23] than the AGA8-DC92 EoS [20,21], with $\Delta_{AAD}$ for the AGA equation that ranges from (0.05 to 0.06) %, in contrast with $\Delta_{AAD}$ for the GERG equation from (0.02 up to 0.04) %. The reason for this relative underperformance of the GERG-2008 EoS [22,23] for hydrogen-containing mixtures could be the non-existent departure function due to the absence of accurate and wide-range experimental data able to model the binary interactions between hydrogen and such important components for natural gas-like mixtures, as nitrogen, carbon dioxide, ethane, propane and butane, and of fitted combining rules with pentane. In addition to the existing development of a binary specific departure function with methane, the EoS are still worthwhile to be improved, so there is a need for consolidated experimental data that can be employed for the building of departure functions, as supported by this paper and discussed elsewhere [41,45].

**4.2 Perfect-gas heat capacities and acoustic virial coefficients.**

Table 8 compares the experimental heat capacity ratios $\gamma^{pg}$ and the molar isobaric heat capacities as perfect-gas $C_{p,m}^{pg}$ with calculated values from the reference AGA8-DC92 EoS [20,21] and the GERG-2008 EoS [22,23]. The hydrogen content of 3 % to a natural gas mixture decreases the $C_{p,m}^{pg}$ between (3.4 up to 4.9) % as the temperature increases.

Both equations of state yield nearly the same heat capacity values as perfect-gas at every temperature and successfully represent the experimental results, with a resulting $\Delta_{AAD}$ = 0.16 % for the 11 M synthetic mixture and a $\Delta_{AAD}$ = 0.27 % for the $H_2$-enriched mixture, respectively, well below the expanded ($k$ = 2) model uncertainty $U_{EoS}(C_{p,m}^{pg})$ = 1.0 %.

However, when comparing with the 11 M synthetic mixture only at the isotherms $T$ = (325 and 350) K, the relative differences are within the mean experimental expanded ($k$ = 2) uncertainty $U_{exp}(C_{p,m}^{pg})$ = 0.085 %. By contrast, with respect to the $H_2$-enriched mixture, the disagreement is almost twofold at all temperatures, always beyond the corresponding mean experimental expanded ($k$ = 2) uncertainty $U_{exp}(C_{p,m}^{pg})$ = 0.061 %.

Table 8 also lists the relative deviations of the experimental second acoustic virial coefficient $\beta_a$ and third acoustic virial coefficient $\gamma_a$ from the estimations of the AGA8-DC92 EoS [20,21] and the GERG-2008 EoS [22,23]. Both models overpredict the experimental results at all temperatures with the exception of the lowest isotherm $T$ = 260 K in the 11 M synthetic mixture.

Regarding $\beta_a$, the relative differences can be as much as halved when compared to the GERG-2008 EoS [22,23] than when compared to the AGA8-DC92 [20,21] for both mixtures, worsening the agreement from an $\Delta_{AAD}$ = 2.5 % in the 11 M synthetic mixture up to $\Delta_{AAD}$ = 5.8 % in the $H_2$-enriched mixture with respect to the GERG-2008 EoS [22,23].

With respect to $\gamma_a$, discrepancies raise one order of magnitude more than those of $\beta_a$, but show a decrease with increasing temperatures. This coefficient seems to be insensitive to hydrogen in the natural gas at the studied concentration. Note that, in any case, the relative deviations are far beyond the corresponding mean experimental expanded ($k$ = 2) uncertainties $U_{exp}(\beta_a)$ = 1.0 % and $U_{exp}(\gamma_a)$ = 2.9 %.

**5. Conclusions.**

We report new accurate experimental speed of sound data $w_{exp}(p,T,\bar{x})$ at pressures between $p$ = (0.1 up to 13) MPa and temperatures $T$ = (260 to 350) K of an 11 component synthetic natural gas mixture and a $H_2$-enriched natural gas with a nominal molar percentage $x_{H_2} = 3$ %. Heat capacity ratio $\gamma^{pg}$, perfect-gas heat capacity $C_{p,m}^{pg}$, second $\beta_a$, and third $\gamma_a$ acoustic virial coefficients have been derived from the speed of sound values.

Taking into account the speed of sound $w_{exp}(p,T,\bar{x})$, this study reveals that: (i) AGA8-DC92 EoS [20,21] performs worse than GERG-2008 EoS [22,23], in agreement with other authors [18]; (ii) all the relative deviations fall within the expanded ($k$ = 2) uncertainty of each model, $U_{AGA8\text{-}DC92}(w)$ = 0.2 % and $U_{GERG\text{-}2008}(w)$ = 0.1 %, respectively; (iii) however, there is a clear increment in the relative differences for the $H_2$-containing natural gas mixture compared to the synthetic natural gas, mainly at low temperatures for the lower pressures, and conversely at high temperatures for the higher pressures, which are in the limit of the model uncertainty for the latter case. Analogous results apply to $C_{p,m}^{pg}$, $\beta_a$, and $\gamma_a$.

Considering that most of the relative discrepancies also are beyond our experimental expanded ($k$ = 2) uncertainty $U_r(w_{exp})$ ~ 0.02 %, $U_{exp}(C_{p,m}^{pg})$ ~ 0.07 %, $U_{exp}(\beta_a)$ = 1.0 % and $U_{exp}(\gamma_a)$ = 2.9 %, we conclude that there is still a need for improvement of the current thermodynamic enhance their performance on hydrogen-containing mixtures. This objective can be approached when further accurate and extensive data sets of binary mixtures of hydrogen with nitrogen, carbon dioxide, and other hydrocarbons apart from methane have become available.

**Acknowledgements**

The authors want to thank for the support to European Regional Development Fund (ERDF)/Spanish Ministry of Science, Innovation and University (Project ENE2017-88474-R) and ERDF/Regional Government of "Castilla y León" (Project VA280P18). DVM is supported by the Spanish Ministry of Science, Innovation and Universities ("Beatriz Galindo Senior" fellowship BEAGAL18/00259).